\pdfoutput=1
\documentclass[]{smule}
\usepackage[round,authoryear]{natbib}
\usepackage{hyperref}
\usepackage[colorinlistoftodos]{todonotes}
\usepackage{colortbl}
\usepackage{nicematrix}
\usepackage{amssymb}
\usepackage{amsmath}
\usepackage{booktabs}
\usepackage{adjustbox}
\usepackage{soul}
\usepackage[inline]{enumitem}
\usepackage{booktabs}
\usepackage{color}
\usepackage{xcolor}
\usepackage{bbding}
\usepackage{listings}
\usepackage{multicol}
\usepackage{xspace}
\usepackage{tablefootnote}
\usepackage{libertinus}

\makeatletter
\AtBeginDocument{

}
\makeatother

\title{StylePitcher: Generating Style-Following and Expressive Pitch Curves for Versatile Singing Tasks}

\author[2\dagger]{Jingyue Huang}
\author[2\dagger]{Qihui Yang}
\author[3\dagger]{Fei Yueh Chen}
\author[2]{Julian McAuley}
\author[1]{Randal Leistikow}
\author[1]{Perry R. Cook}
\author[1]{Yongyi Zang}

\affiliation[1]{Smule Labs}
\affiliation[2]{University of California, San Diego}
\affiliation[3]{University of Rochester}

\contribution[\dagger]{Work done during internship at Smule}

\abstract{
    Existing pitch curve generators face two main challenges: they often neglect singer-specific expressiveness, reducing their ability to capture individual singing styles. 
    And they are typically developed as auxiliary modules for specific tasks such as pitch correction, singing voice synthesis, or voice conversion, which restricts their generalization capability. 
    We propose StylePitcher, a general-purpose pitch curve generator that learns singer style from reference audio while preserving alignment with the intended melody. 
    Built upon a rectified flow matching architecture, StylePitcher flexibly incorporates symbolic music scores and pitch context as conditions for generation, and can seamlessly adapt to diverse singing tasks without retraining. 
    Objective and subjective evaluations across various singing tasks demonstrate that StylePitcher improves style similarity and audio quality while maintaining pitch accuracy comparable to task-specific baselines.
}

\begin{document}

\maketitle

\section{Introduction}
Pitch curves, or fundamental frequency (F0) curves, are the backbone of expressive singing. They encode not only the melody but also the subtle variations that define unique styles of different singers, such as their vibrato, ornaments, pitch bending, and others~\citep{SingStyle111}. 
Therefore, pitch curves serve as critical intermediate representations across diverse singing generation and conversion tasks, such as automatic pitch correction (APC)~\citep{TargetAcquisition, KaraTuner, Diff-Pitcher}, singing voice synthesis (SVS)~\citep{DiffSinger, VISinger, RMSSinger, ExpressiveSinger, StyleSinger,  TCSinger}, and singing voice conversion (SVC)~\citep{SVC-Non-parallel, Pitchnet, ImprovingAdversarial, AHierarchicalSpeaker, UCD-SVC, SPA-SVC, SYKI-SVC}.

Despite their importance, existing approaches face two main limitations. 
First, most of them overlook singer-specific styles encoded in pitch curves, treating pitch as a singer-agnostic feature and reusing the same curve across different singers~\citep{SVC-Non-parallel, Pitchnet}.
This limitation is critical: the same melody sung by different singers can produce distinct pitch patterns, which reflect individual singing techniques and styles. 
Losing these patterns will neglect their singer-specific expressiveness and the essence of their performance~\citep{ExpressiveSinger}. 
Second, while some recent approaches~\citep{Diff-Pitcher, StyleSinger} support style-informed pitch curve generation, they often develop the pitch curve generator as an auxiliary module for specific tasks, such as pitch correction~\citep{Diff-Pitcher} and singing voice synthesis~\citep{StyleSinger}. 
This task-specific design constrains their generalization capability across different singing applications, as researchers have to retrain these modules with different hyperparameters, inputs and outputs for each adaptation. 
%
Therefore, a general-purpose model capable of generating style-following pitch curves is essential for diverse singing applications.

We propose \textbf{StylePitcher}, the first style-following pitch curve generation model for versatile singing tasks.
We formulate pitch curve generation as a masked infilling problem: given surrounding pitch context and symbolic music scores, StylePitcher learns to generate missing pitch segments that continue the style patterns from context.
This approach enables implicit style modeling without requiring explicit singer labels or embeddings, allowing generalization to unseen voices.
We employ a rectified flow model~\citep{RFM} for stable, efficient, and high-quality generation process. 
In addition, we introduce a smoothing algorithm to construct reliable conditioning signals (i.e., symbolic music scores), removing the need for manual annotations. 
By separately modeling F0 and performing inpainting, StylePitcher generates pitch curves that follow the style of provided audio without any task-specific retraining.
Once trained, it serves as a plug-and-play module for diverse applications, including pitch correction, zero-shot singing voice synthesis with style transfer and style-informed singing voice conversion. 

\begin{figure*}[t]
 \centerline{
 \includegraphics[width=\textwidth]{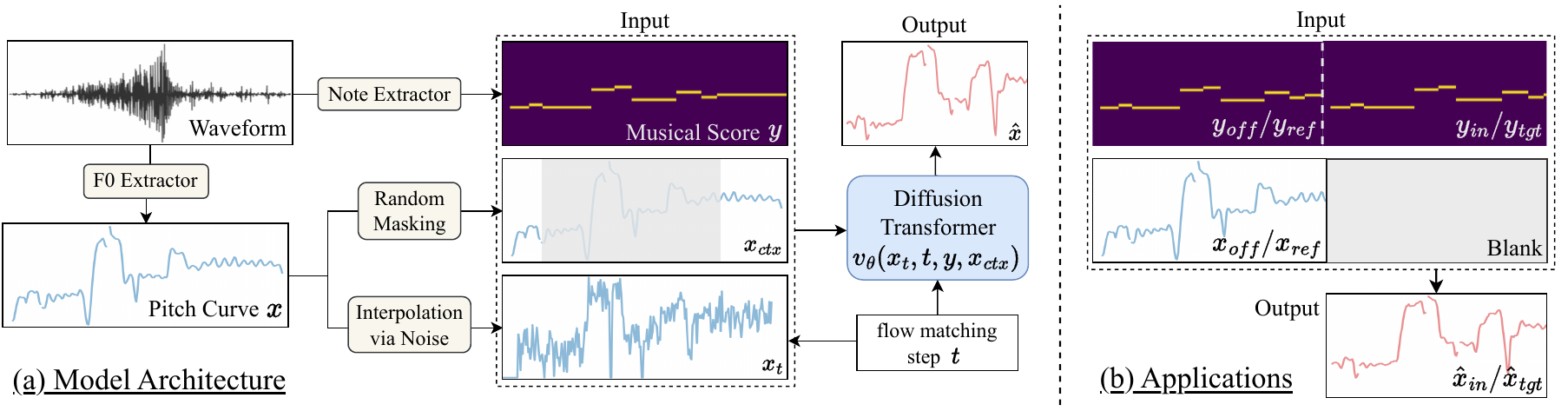}}
 \vspace{-0.25cm}
 \caption{Illustration of the methods. The unvoiced condition is omitted for clarity. Subscripts \texttt{off} and \texttt{in} denote features from off-key and in-key singing in the APC task; \texttt{ref} and \texttt{tgt} refer to reference and target content for SVS and SVC tasks.}
 \vspace{-0.2cm}
 \label{fig:model}
\end{figure*}

Our contributions are threefold:
\begin{itemize}[leftmargin=*, itemsep=0pt, topsep=2pt]
    \item StylePitcher\footnote{Demos available at \url{https://stylepitcher.github.io/}} is the first general-purpose, style-following pitch curve generator supporting diverse singing tasks.
    \item We introduce a flow matching architecture to pitch curve generation, a smoothing algorithm for data annotation, and an inpainting mechanism for flexible task adaptation. 
    \item Objective and subjective evaluations across multiple singing tasks show that StylePitcher achieves superior or compatible performance on style similarity, audio quality and pitch accuracy relative to previous baselines.
\end{itemize}

\section{Related Work}

Many singing generation frameworks explicitly predict pitch curves as an intermediate step. 
In automatic pitch correction, models generate pitch contours conditioned on target note sequences to correct out-of-tune vocals~\citep{TargetAcquisition, KaraTuner, Diff-Pitcher}.
Singing voice synthesis aims to generate singing voice from lyrics and scores, with many approaches predicting F0 to model pitch variation~\citep{DiffSinger, VISinger, RMSSinger, ExpressiveSinger, TCSinger}. 
Specifically, StyleSinger~\citep{StyleSinger} adopts style-specific and style-agnostic pitch predictors to capture singing styles from references.
Singing voice conversion models transform the voice in a singing signal into that of a target singer while preserving content and melody. 
Most existing works~\citep{SVC-Non-parallel, Pitchnet, UCD-SVC, AHierarchicalSpeaker, ImprovingAdversarial, SPA-SVC, SYKI-SVC} treat pitch as singer-agnostic, reusing the unchanged or key-shifted F0 sequences from the original singing, which converts timbre but overlooks singer-specific expressiveness.
Unlike prior works that embed task-specific F0 predictors within complex frameworks, the proposed StylePitcher is a plug-and-play module for diverse singing tasks, supporting style preservation or transfer as needed.

\section{Methods}
\vspace{-0.15cm}
\subsection{Rectified Flow}

Rectified flow~\citep{RFM} defines generative modeling as learning a deterministic transport map from a prior distribution $\pi_0$ (e.g. Gaussian) to the data distribution $\pi_1$.
It parameterizes a time-dependent velocity field $v_\theta(x_t, t, c)$ to push $x_0 \sim \pi_0$ towards $x_1 \sim \pi_1$ under the ordinary differential equation (ODE):
\vspace{-0.1cm}
\begin{equation}
    d x_t = v_\theta(x_t, t, c)dt 
    \vspace{-0.1cm}
\end{equation}
where $c$ denotes conditions, $t\in [0, 1]$ is the scalar time parameter, and $x_t$ follows a rectified linear interpolation:
\vspace{-0.1cm}
\begin{equation}
    x_t = (1-t) x_0 + t x_1.
    \vspace{-0.1cm}
\end{equation}
The model is trained with the flow matching objective:
\vspace{-0.1cm}
\begin{equation}\label{eq:objective}
    \mathcal{L}(\theta) = \mathbb{E}_{x_0 \sim \pi_0, x_1 \sim \pi_1, t, c}\| v_\theta(x_t, t, c) - (x_1 - x_0)\|^2_2.
    \vspace{-0.1cm}
\end{equation}
Once trained, starting from 
$\pi_0$, samples from $\pi_1$ are obtained by integrating the learned ODE from $t=0$ to $t=1$, producing high-quality results with few sampling steps.

\subsection{Model Architecture}
As illustrated in Figure \ref{fig:model}, we formulate pitch curve generation as a conditional infilling task, following Voicebox~\citep{Voicebox}. 
Given a fundamental frequency curve $x = (x^1, \cdots, x^N)$ from a singing voice, the corresponding note sequence $y = (y^1, \cdots, y^N)$, and a binary mask $m \in \{0, 1\}^N$, our model $p(x_{\text{mask}} | y, x_{\text{ctx}})$ predicts the masked segments $x_{\text{mask}} = m \odot x$ conditioned on the complete note sequence $y$ and the context $x_{\text{ctx}} = (1-m) \odot x$.
Through in-context learning, the generated segments implicitly follow the singing style of the surrounding context remaining aligned with the target musical score. 

We adopt rectified flow with a diffusion transformer~\citep{Transformer, peebles2023scalable} to parameterize the velocity field $v_\theta(x_t, t, y, x_{\text{ctx}})$, where the full signal $x$ is modeled instead of $x_{\text{mask}}$ for simpler conditioning, and $x_t$ is the linear interpolation between noise $\epsilon \sim \mathcal{N}(0, 1)$ and $x$ at flow matching step $t$. 

Specifically, the pitch curves $x_t$ and $x_{\text{ctx}}$ are linearly projected to embeddings of shape $(N,H_1=512)$, and notes $y \in [M]^N$ are projected to $(N,H_2=256)$, where $M$ is the number of pitch classes.
These three embeddings are concatenated along the frame dimension and projected to the embeddings for next layers. Additionally, an unvoiced indicator sequence $u \in \{0, 1\}^N$ is incorporated to align the generated F0 with singing phonemes.
Finally, a sinusoidally positional-encoded flow step $t$ modulates the representation to form the transformer input. 

The training objective is:
\begin{equation}
    \mathcal{L}_{pitch}(\theta) = \mathbb{E}_{\epsilon, p(x, y), t, m}\|m \odot \left[v_\theta(x_t, t, y, x_{\text{ctx}}) - (x - \epsilon)\right]\|^2_2
\end{equation}
where the loss is computed only on masked frames~\citep{Voicebox}.
Classifier-free guidance (CFG)~\citep{CFG} is employed by randomly dropping conditions $y$, $x_{ctx}$ and $u$ with probability $p_c$ for training.  
During inference, samples are generated by integrating the ODE for $K$ steps with modified velocity field $\hat{v}_\theta$ and CFG scale $\alpha$:
\begin{equation}
    \hat{v}_\theta = v_\theta(x_t, t, \varnothing) + \alpha [v_\theta(x_t, t, y, x_{\text{ctx}}) - v_\theta(x_t, t, \varnothing)].
\end{equation}

\subsection{Applications}
As a standalone model, StylePitcher performs task-agnostic pitch curve generation via conditional inpainting, supporting style preservation or transfer across diverse singing applications without the need for re-training, as shown in Figure \ref{fig:model}(b).

\vspace{0.22cm}
\noindent\textbf{Automatic Pitch Correction}\; Given off-key singing with F0 $x_{\text{off}}$, notes $y_{\text{off}}$, unvoiced sequence $u_{\text{off}}$, and target notes $y_{\text{in}}$, we construct the input as $x = (x_{\text{off}}, 0)$ and generate $\hat{x}$ conditioned on $y = (y_{\text{off}}, y_{\text{in}})$ and $u = (u_{\text{off}}, u_{\text{off}})$.
The corrected pitch $\hat{x}_{\text{in}}$ is obtained from the latter portion of $\hat{x}$, preserving the original singing style and matching the target notes.

\vspace{0.22cm}
\noindent\textbf{Zero-Shot SVS with Style Transfer}\; Given reference singing ($x_{\text{ref}}$, $y_{\text{ref}}$, $u_{\text{ref}}$) and target content from an SVS model ($x_{\text{tgt}}$, $y_{\text{tgt}}$, $u_{\text{tgt}}$), we concatenate the three sequences and mask the target segment $x_{\text{tgt}}$. 
StylePitcher then generates $\hat{x}_{\text{tgt}}$ that aligns with the target notes while following the reference style, which replaces $x_{\text{tgt}}$ for SVS synthesis.

\vspace{0.22cm}
\noindent\textbf{Style-informed SVC}\; Unlike previous SVC models using unchanged F0, we modify pitch contours to capture singing style. 
Given reference and target audio features concatenated as $x = (x_{\text{ref}}, x_{\text{tgt}})$ with corresponding $y$ and $u$, we mask and regenerate $x_{\text{tgt}}$ to obtain $\hat{x}_{\text{tgt}}$, enabling the converted singing to transfer both timbre and pitch style while preserving content.

\vspace{-4pt}
\section{Experiments}

\subsection{Datasets and Data Processing}\label{sec:data}

For training, we employ two multi-speaker singing datasets, DAMP-VSEP~\citep{DAMP-VSEP} and DAMP-VPB~\citep{DAMP-VPB}, totaling 1916 hours of singing voice. 
For evaluation, different test sets are used depending on the task: (1) Samples from Diff-Pitcher~\citep{Diff-Pitcher} for pitch correction; (2) GTSinger~\citep{GTSinger} for singing voice synthesis and conversion; (3) VocalSet~\citep{VocalSet} for technique diversity.

We adopt RMVPE~\citep{RMVPE} for F0 estimation and unvoiced detection (16 kHz, 1024 frame size, 160 hop size); Basic Pitch~\citep{Basic-Pitch} for MIDI extraction. Empirically, we replace the multi-pitch activation of Basic Pitch with that of RMVPE to obtain more accurate MIDI extraction results.
The output F0 spans \textit{C1} (32.7 Hz) to \textit{B6} (1975.5 Hz), covering $M=72$ note classes. 
Audio is pre-processed with vocal separation and denoising before F0 extraction.
We observe that the extracted MIDI still contains style information expressed as short notes. 
To remedy, we apply Gaussian blur on the multi-pitch activation map to smooth expressive techniques (e.g., vibrato) and post-process by removing short rests and notes.


\begin{table}[t]
\small\centering
\begin{tabular}{l|ccc|c}
\toprule
\textbf{Models}         & \textbf{RPA} $\uparrow$   & \textbf{PCA} $\uparrow$   & \textbf{OA} $\uparrow$    & \textbf{Acc.} $\downarrow$ \\ 
\midrule
Diff-Pitcher            & 67.37                 & 67.40                 & 70.30                 & 69.43                  \\
StyleSinger             & -                     & -                     & -                     & 71.48                  \\
\midrule
StylePitcher              & \underline{68.64}     & \underline{68.74}     & \underline{73.04}     & \textbf{51.85}         \\ 
\quad -- w/o smo. & \textbf{69.49}        & \textbf{69.61}        & \textbf{73.61}        & 52.71                   \\
\quad -- w/o ctx.      & 66.71                 & 66.82                 & 71.34                 & \underline{52.12}       \\
\bottomrule
\end{tabular}
\vspace{-2pt}
\caption{Objective evaluations (\%) on GTSinger dataset.}
\vspace{-10pt}
\label{table:objective}
\end{table}

\subsection{Experimental Setting}
We use an 8-layer, 8-head diffusion transformer (DiT) with 512 hidden dim. and rotary position embeddings~\citep{RoFormer}, totaling 49M parameters. 
The maximum sequence length is $N$=$1024$ frames (20.48 seconds at 50 Hz). 
We apply cosine schedule~\citep{CosMap} to focus on lower $t$ values, mask $r\%$ of sequences for infilling ($r \sim \mathcal{U}[70, 100]$), and set CFG drop probability $p_c=0.5$. 
Pitch curves and notes are augmented by random shifts within $[-4, 4]$ semitones. 
The model is pretrained for 100k steps without unvoiced conditioning (learning rate 1e-4) and fine-tuned for 90k steps with it (1e-5), using 5k-step linear warm-up, AdamW~\citep{AdamW} optimizer, cosine scheduler, and batch size 512. 
Training is done with FlashAttention-2~\citep{FlashAttention-2}. 
During inference, we use the midpoint solver of torchdiffeq~\citep{torchdiffeq} with $K$=$16$ steps and CFG scale $\alpha$=$1.25$. 
Generated F0 curves can be interpolated to match the expected F0 sampling rates for downstream singing tasks.

\begin{table*}[t]
\centering
\scalebox{1}{
\begin{tabular}{l|ccc|cc|cc}
\toprule
\multirow{2}{*}{\textbf{Models}} & \multicolumn{3}{c|}{\textbf{APC}} & \multicolumn{2}{c|}{\textbf{SVS}} & \multicolumn{2}{c}{\textbf{SVC}} \\
                        & MOS-P     & MOS-S     & MOS-Q     & MOS-S             & MOS-Q        & MOS-S          & MOS-Q           \\ 
\midrule
\multirow{2}{*}{*Baselines} & \multicolumn{3}{c|}{Diff-Pitcher~\citep{Diff-Pitcher}} & \multicolumn{2}{c|}{StyleSinger~\citep{StyleSinger}} & \multicolumn{2}{c}{In-house SVC} \\
             & \textbf{4.18}\footnotesize{$\pm$0.21}  & 3.38\footnotesize{$\pm$0.20} & \underline{3.09}\footnotesize{$\pm$0.18} & 3.21\footnotesize{$\pm$0.22} & 3.07\footnotesize{$\pm$0.19}  & 2.62\footnotesize{$\pm$0.23} & \textbf{3.03}\footnotesize{$\pm$0.22}\\
\midrule
StylePitcher              & \underline{3.84}\footnotesize{$\pm$0.22} & \textbf{3.64}\footnotesize{$\pm$0.20} & \textbf{3.26}\footnotesize{$\pm$0.18} & \underline{3.33}\footnotesize{$\pm$0.23} & \underline{3.11}\footnotesize{$\pm$0.23} & \textbf{2.95}\footnotesize{$\pm$0.25} & \underline{2.72}\footnotesize{$\pm$0.22} \\ 
\quad -- w/o smo.       & 3.66\footnotesize{$\pm$0.24}  & \underline{3.39}\footnotesize{$\pm$0.19} & 3.04\footnotesize{$\pm$0.19} & \textbf{3.45}\footnotesize{$\pm$0.21} & \textbf{3.18}\footnotesize{$\pm$0.21} & \underline{2.72}\footnotesize{$\pm$0.22} & 2.64\footnotesize{$\pm$0.23} \\
\bottomrule
\end{tabular}
}
\caption{Subjective evaluations on three singing tasks. *Baselines correspond to Diff-Pitcher, StyleSinger, and the in-house SVC for their respective tasks. MOS-P, MOS-S, and MOS-Q refer to mean opinion scores for Pitch, Style, and Quality aspects.} 
\vspace{-2pt}
\label{table:subjective}
\end{table*}

\subsection{Baselines and Metrics}
We compare against three baselines to evaluate generated pitch curves under task-specific settings, focusing on style capture ability: (1) Diff-Pitcher~\citep{Diff-Pitcher} for APC; (2) StyleSinger~\citep{StyleSinger} for zero-shot SVS with style transfer; (3) an in-house SVC model using unchanged F0.
%
We evaluate only their pitch prediction modules where applicable. 
Two ablations are included to assess the proposed smoothing algorithm (w/o smo.) and the inpainting setting (w/o ctx., with mask $m=0$).

For objective metrics, we measure pitch alignment \footnote{StyleSinger is excluded because annotated scores are not perfectly aligned with the audio.} 
using Raw Pitch Accuracy (RPA, within half-semitone), Raw Chroma Accuracy (RCA, octave-invariant RPA), and Overall Accuracy (OA, including non-melody frames)~\citep{mcfee2025librosa}.
To assess overall similarity, we train a 2-layer LSTM model on the curves and report classification accuracy (Acc.) between generated and ground-truth pitch, where \textbf{lower values indicate higher similarity}. 
These metrics are evaluated on the Chinese GTSinger set~\citep{GTSinger}, unseen by all compared models.

For subjective evaluation, we conduct an online listening test on all three tasks. Participants first listened to the off-key voice (APC) or reference singing tracks (SVS/SVC), then rated generated samples from different models on 5-point Likert scales in three aspects: (1) pitch correction accuracy (APC only); (2) style preservation (APC/SVS) or capture (SVC); (3) overall quality. 
For SVS, we evaluate only cases where reference content remains unchanged. 
%
%
We collected 19 responses from participants with diverse musical backgrounds, yielding 76 ratings per task per model per aspect.

\section{Results and Discussions}

\subsection{Objective Evaluation}

Table \ref{table:objective} shows that StylePitcher outperforms all baselines across pitch alignment and similarity metrics. 
Notably, the LSTM classifier achieves near-random accuracy (50\%) when distinguishing our generated curves from real ones, demonstrating the effectiveness of rectified flow for modeling continuous signal. 
The ablation without smoothing achieves slightly better alignment metrics by adhering more strictly to musical scores, while removing context degrades performance, confirming the benefit of in-context learning.

\subsection{Subjective Evaluation}

Table \ref{table:subjective} presents human evaluation results across three tasks.

\begin{figure}[t]
\centering
\includegraphics[width=0.83\linewidth]{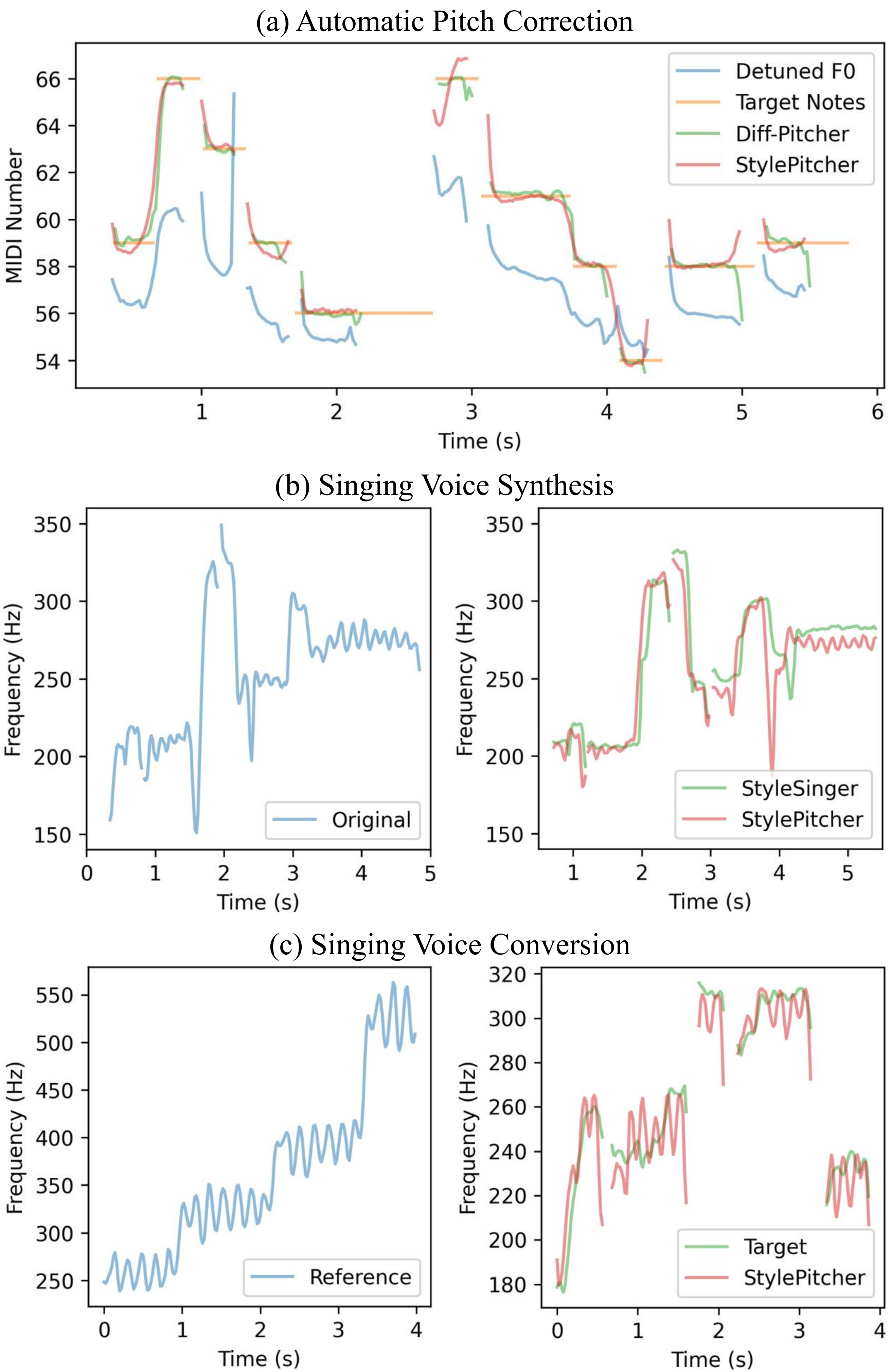}
\vspace{-5pt}
\caption{Samples for three singing tasks. StylePitcher (red) captures singing styles better from input curves (blue) than baselines (green), such as pitch slides (a) and vibrato (b\&c).}
\label{fig:example}
\vspace{-8pt}
\end{figure}

\vspace{0.2cm}
\noindent\textbf{Automatic Pitch Correction}\;
StylePitcher better preserves singing style and audio quality than Diff-Pitcher~\citep{Diff-Pitcher}, though with lower pitch correction accuracy. 
As shown in Figure \ref{fig:example}(a), our method maintains expressive elements like pitch slides while correcting notes, producing more personalized corrections rather than enforcing strict alignment.

\vspace{0.2cm}
\noindent\textbf{Zero-Shot SVS with Style Transfer}\; 
Despite never training jointly with synthesis frameworks, StylePitcher achieves superior style capture compared to StyleSinger~\citep{StyleSinger} and maintaining comparable audio quality. 
Figure \ref{fig:example}(b) also demonstrates that our method effectively captures vibrato and glissando characteristics that StyleSinger misses, validating its potential as a plug-and-play module for enhanced expressiveness.

\vspace{0.2cm}
\noindent\textbf{Style-informed SVC}\; 
Unlike the baseline using unchanged F0, StylePitcher successfully transfers both timbre and singing style. 
Figure \ref{fig:example}(c) shows the transformation of a flat target curve into one with strong vibrato from the reference. 
However, applying expressive techniques without content awareness can occasionally produce unnatural results, impacting audio quality. 
We leave resolving this limitation to future work.

\vspace{0.2cm}
These results validate StylePitcher as an effective general-purpose pitch generator that balances pitch accuracy with style capture across diverse tasks, enabling expressive singing applications without task-specific training.

\section{Conclusion}

We presented StylePitcher, a general-purpose pitch generation framework that captures and transfers singing styles through masked infilling with rectified flow. Without task-specific training or manual annotations, our DiT-based model achieves superior performance across automatic pitch correction, singing voice synthesis, and voice conversion. Its plug-and-play design enables immediate deployment in existing systems. Future work will explore content-aware generation and extend to other performance parameters for comprehensive style modeling.

\clearpage
\newpage
\bibliographystyle{plainnat}
\bibliography{paper}

\end{document}